\documentclass[aps,preprint,nofootinbib]{revtex4}

\usepackage{graphicx}
\usepackage{amssymb}
\usepackage{amsmath}
\usepackage{epstopdf}
\usepackage{ulem}

\usepackage[greek,english]{babel}
\usepackage[iso-8859-7]{inputenc}

\usepackage{color}

\def\ba{\begin{eqnarray}}
\def\ea{\end{eqnarray}}
\def\be{\begin{equation}}
\def\ee{\end{equation}}

\def\gtorder{\mathrel{\raise.3ex\hbox{$>$}\mkern-14mu
             \lower0.6ex\hbox{$\sim$}}}
\def\ltorder{\mathrel{\raise.3ex\hbox{$<$}\mkern-14mu
             \lower0.6ex\hbox{$\sim$}}}

\def\dalemb#1#2{{\vbox{\hrule height.#2pt
  \hbox{\vrule width.#2pt height#1pt \kern#1pt \vrule width.#2pt}
    \hrule height.#2pt}}}

\def\fwhm{\theta_{\rm{fwhm}}}
\def\sigpix{\sigma_{\rm{pix}}}
\def\fsky{f_{\rm{sky}}}
\def\ns{n_{\rm s}}

\newcommand{\planck}{{\sc Planck}} 
\newcommand{\designer}{{\sc Designer}}

\begin{document}
\rightline{T$\Theta\Delta$}

\title{CMB lensing reconstruction with point source masks}
\author{C.~Sofia~Carvalho${}^{1,2}$\footnote{Email: carvalho.c@gmail.com} 
and
Ismael~Tereno${}^{3}$\footnote{Email: tereno@fc.ul.pt}}
\address{$^1$ 
Instituto de Plasmas e Fus\~ao Nuclear, Instituto Superior T\'ecnico,
Avenida Rovisco Pais, 1, 1049-001 Lisboa, Portugal}
\address{$^2$
Academy of Athens, Research Center for Astronomy and Applied Mathematics, Soranou Efessiou 4, 11-527, Athens, Greece}
\address{$^3$ Centro de Astronomia e Astrof\'{i}sica da Universidade de Lisboa, Tapada da Ajuda, 1349-018 Lisboa, Portugal}
\begin{abstract}
An incomplete sky coverage poses difficulties in the extraction of the weak lensing information from the CMB.
We test the reconstruction of the weak lensing convergence from CMB maps to which masks of point sources have been applied. We use the quadratic estimator with a kernel with finite support acting in real space for a \planck~simulation. We recover the lensing signal without significant loss of power or addition of spurious correlations, thus showing that masking defected pixels does not affect the reconstruction of the weak lensing convergence in real space.
\end{abstract}
\date{\today}
\maketitle

\section{Introduction}

The weak gravitational lensing of the CMB
is a measure of the gravitational potential projected along the line
of sight from the last scattering surface. It was indirectly detected using WMAP 
data \cite{ksmith07,hirata08} and recently direct detections in WMAP 
\cite{smidt10} and ACT \cite{das11} were reported.

In a CMB experiment, the temperature signal is mixed with various
foreground signals which can generate spurious correlations.
To obtain the signal of interest, we must first remove the
contaminating signals. However, an incorrect removal could create new
correlations.
In order to robustly test estimators of quantities measured from the
CMB, we need to ensure a correct removal of the contaminants without
degrading the signal of interest.

Any process resulting from the coupling of different Fourier modes is
sensitive to contaminants. The weak gravitational lensing of the CMB induces 
non-Gaussianities
in the form of mode couplings and consequently is very sensitive to
spurious correlations from foreground signals.
The contaminants can be either physical components (integrate Sachs-
Wolfe effect, Sunyaev-Zel'dovich effect, galactic foregrounds, point
sources) or experimental components (instrumental noise, beam effects).
Such systematics have become more relevant with the current generation of 
high-precision CMB experiments such as ACT {\footnote {\tt http://www.physics.princeton.edu/act/about.html}}, \planck~{\footnote {\tt http://www.rssd.esa.int/index.php?project=planck}}  or SPT {\footnote {\tt http://pole.uchicago.edu/}}. 
Here we are concerned with the treatment of point sources on CMB maps.

Point sources can be removed either by subtracting the estimated
emission, through a model of the point source flux, or by discarding
the contaminated pixels above a certain flux.
Filtering the emission from point sources was the method used in the
recent detection of the convergence power spectrum from the ACT data \cite{das11}. 
Another approach is called inpainting and consists in removing the point sources and replacing them with an extrapolation of the
missing data using some prior of the complete data
\cite{inpaint,inpaint2}.
However, both the subtraction of the point source emission and the
inpainting of masking disks are dependent on the extrapolation
model (i.e. the model of the point source flux and the prior of the
complete sky, respectively). To extract a model independent result, we
should be able to simply discard the contaminated pixels by applying a
point source mask to the map we want to analyse.

Discarding the pixels amounts to multiplying the map by a mask, which
causes spatial modulation of the average sky brightness and
consequently affects the analysis of CMB experiments.
The multiplication of the map by a mask for all point sources can be
regarded as the successive multiplication of the map by a mask for
each point source \cite{taper}.
The Fourier transform of the masked map will be equivalent to
successively convolving the Fourier transform of the original map by
that of a series of such masks, which will mix power of the map over a
range of multipoles.
The reconstructed power spectrum must then be carefully corrected from these 
contaminants introduced by the analysis of incomplete data. 


Here we want to study to which extent masking small areas at the
locations of point sources affects the weak lensing
convergence extracted from CMB maps with the estimator introduced in
Ref.~\cite{carvalho}. This 
is a variant of the quadratic estimator with the novelty of having a
finite support and acting in real space. (See also Ref.~\cite{bucher}
for a study of its properties.) We also compare the
results with those obtained from the conventional quadratic estimator
defined in harmonic space.

The manuscript is organized as follows. In section \ref{sec:estimator}
we describe how we estimate the lensing convergence using the minimum
variance estimator and how we deal with the bias of the estimator. In
section \ref{sec:application} we synthesise the point source masks and
apply the estimator to lensed CMB maps with and without masking.
In section \ref{sec:degradation} we study the degradation of the
estimation of the convergence power spectrum as a function of the
parameters of the mask, namely the point source number density and the
masking disk diameter per point source.
We summarize our results in section \ref{sec:summary}.

\section{Estimator of the weak lensing convergence}
\label{sec:estimator}

Weak lensing of the 
CMB consists in the deflection of CMB photons from the original propagation direction $\boldsymbol{\theta}$ on the last scattering surface (the source plane) to an observed direction $\tilde{\boldsymbol{\theta}}$ on the sky today (the image plane) (see Ref.~\cite{lewis&challinor06} for a review).
This deflection amounts to remapping the unlensed temperature anisotropies $T$ to the lensed ones $\tilde T$ according to 
\ba
\tilde T(\boldsymbol{\theta})
=T(\tilde{\boldsymbol{\theta}})=
T(\boldsymbol{\theta}+\nabla \psi)
=T(\boldsymbol{\theta})
+\nabla\psi \cdot \nabla T(\boldsymbol{\theta})
+O[(\nabla\psi)^2].
\label{eq:cmblensing}
\ea
The deflection angle 
${\boldsymbol{\alpha}}= \tilde{\boldsymbol{\theta}}-\boldsymbol{\theta}$
is given by ${\boldsymbol{\alpha}}=\nabla\psi,$
where the lensing potential $\psi$ is the projection, on the image plane and along the line-of-sight, of the gravitational potential of the intervening large-scale structure, and $\nabla$ is the covariant derivative on the image plane.
The right-hand-side of Eq.~(\ref{eq:cmblensing}) is written for small deflections as a first-order Taylor expansion in the deflection, showing that the lensed CMB temperature depends on the derivative of the unlensed temperature through the coupling with the deflection.   

For small deflections, the lensing effect is described by the amplification matrix $\boldsymbol{A}=1-\boldsymbol{\kappa}$, where the convergence tensor $\boldsymbol{\kappa}$ is defined by second-order derivatives of the projected potential. In the absence of rotations, $\boldsymbol{\kappa}$
 can be decomposed as a sum of an isotropic (diagonal) and an anisotropic (traceless and symmetric) part as follows
\ba
\boldsymbol{\kappa}
=\left(
\begin{array}{cc}
\kappa_{0}+\kappa_{+} & \kappa_{\times}\\
\kappa_{\times} & \kappa_{0}-\kappa_{+}
\end{array}
\right).
\ea
The isotropic part produces a convergence/dilation whereas the anisotropic part produces a shear. In particular, the convergence $\kappa_{0}=-(\partial_{x}^2+\partial_{y}^2)\psi/2$ magnifies a feature on the last-scattering surface, the shear $\kappa_{+}=-(\partial_{x}^2-\partial_{y}^2)\psi/2$ stretches it along the $x$--axis while compressing it along the $y$--axis, and analogously the shear $\kappa_{\times}=-\partial_{x}\partial_{y}\psi$ stretches it along the $y=x$ axis and compresses it along the $y=-x$ axis.

From Eq.~(\ref{eq:cmblensing}) the lensed temperature power spectrum writes
\ba
\left< \tilde T(\boldsymbol{\ell}^{\prime})~
\tilde T(\boldsymbol{\ell}-\boldsymbol{\ell}^{\prime})\right>
=(2\pi)^2~
\delta_{\rm D}(\boldsymbol{\ell})~C_{\ell^{\prime}}
+(2\pi)^2
\left[ \boldsymbol{\ell}\cdot\boldsymbol{\ell}^{\prime}~
C_{\ell^{\prime}}
+\boldsymbol{\ell}\cdot(\boldsymbol{\ell}-\boldsymbol{\ell}^{\prime})~
C_{\vert \boldsymbol{\ell}-\boldsymbol{\ell}^{\prime}\vert}\right]
\psi(\boldsymbol{\ell}).
\label{eqn:c_tt}
\ea
The lensing field can thus be reconstructed with a quadratic estimator \cite{hu01}, i.e. a weighted convolution of the square of the lensed map $\tilde T(\boldsymbol{\ell})$. The lensing information is contained in the off-diagonal terms of the lensed $\tilde T(\ell) \tilde T(\ell')$ correlation that are generated by anisotropies induced by the lensing potential. In Ref.~\cite{okamoto&hu03} an optimal weighting is derived as well as a normalization that, in the assumption of Gaussian temperature and lensing random fields, and Gaussian noise uncorrelated with the signal, ensures the estimator has no calibration bias.

Here we will use the real-space estimator introduced in Ref.~\cite{carvalho}, which is a real-space analog of the optimal estimator based on the convolution of the square of the lensed temperature map with a local kernel as follows
\ba
\hat \kappa_{0}(\boldsymbol{\theta})
=\int d^2\boldsymbol{\theta}^{\prime}~\tilde T(\boldsymbol{\theta}^{\prime})
\int d^2\boldsymbol{\theta}^{\prime\prime}~\tilde T(\boldsymbol{\theta}^{\prime\prime})~
Q(\boldsymbol{\theta},\boldsymbol{\theta}^{\prime},\boldsymbol{\theta}^{\prime\prime}).
\ea
Here the kernel $Q(\boldsymbol{\theta},\boldsymbol{\theta}^{\prime},\boldsymbol{\theta}^{\prime\prime})$ is related to that of the standard quadratic estimator $Q(\boldsymbol{\ell},\boldsymbol{\ell}^{\prime})$ by
\ba
Q(\boldsymbol{\theta},\boldsymbol{\theta}^{\prime},
\boldsymbol{\theta}^{\prime\prime})
=\int {d^2\boldsymbol{\ell}\over {(2\pi)^2}}~
\exp[i\boldsymbol{\ell}\cdot\boldsymbol{\theta}]
\int {d^2\boldsymbol{\ell}^{\prime}\over {(2\pi)^2}}~
\exp[-i\boldsymbol{\ell}^{\prime}\cdot\boldsymbol{\theta}^{\prime}]
\exp[-i(\boldsymbol{\ell}-\boldsymbol{\ell}^{\prime})
\cdot\boldsymbol{\theta}^{\prime\prime}]~
Q(\boldsymbol{\ell},\boldsymbol{\ell}^{\prime}).
\ea
The kernel $Q(\boldsymbol{\ell},\boldsymbol{\ell}^{\prime})$ is the minimum variance (to leading order) weight function that convolves the temperature map in the estimation of the weak lensing convergence in harmonic space
\ba
\hat \kappa_{0}(\boldsymbol{\ell})
=\ell^2\int {d^2\boldsymbol{\ell}^{\prime}\over {(2\pi)^2}}~
\tilde T(\boldsymbol{\ell}^{\prime})~
\tilde T(\boldsymbol{\ell}-\boldsymbol{\ell}^{\prime})~
Q(\boldsymbol{\ell},\boldsymbol{\ell}^{\prime}).
\ea
For comparison of the results, we will also use the harmonic-space estimator.
The estimators will be applied to lensed CMB maps synthesised in the flat-sky approximation. 

\subsection{The estimated map}
The reconstructed convergence maps contain not only the contribution due to the lensing potential but also a contribution due to the unlensed CMB temperature, and we write it schematically as
\ba
\hat \kappa_{0}=\kappa_{0\vert\psi}+\kappa_{0\vert\psi=0}.
\label{eq:estimatorsimples}
\ea

The non-lensing contribution is due to other anisotropic couplings in the CMB temperature field, such as $T\,\nabla T$ correlations. These vanish when taking the average over realizations of the CMB (i.e. realizations of lensed CMB maps with the same lensing potential, being thus a source of noise), but are not zero in a given realization.
A convergence map reconstructed from one unlensed CMB realization has a rms per pixel of $\sigma_{\kappa_{0\vert\psi=0}}=0.1$. This is the same amplitude as the rms of a convergence map reconstructed from a lensed CMB map, showing that the lensing contribution is sub-dominant. Over 100 realizations, the rms reduces to $\sigma_{\kappa_{0\vert\psi=0}}=0.01.$
Hence the estimator is noisy but unbiased.

However, the estimator is biased if some assumptions are dropped. Firstly, if all configurations of $\boldsymbol{\ell}$, $\boldsymbol{\ell}^{\prime}$ and $(\boldsymbol{\ell}-\boldsymbol{\ell}^{\prime})$ are used when computing Eq.~(\ref{eqn:c_tt}), the diagonal terms of the lensed $\tilde T(\ell) \tilde T(\ell')$ correlation will also contribute to $\kappa_{0\vert\psi=0}$ by introducing an additive bias in the estimator. Since both $\kappa_{0\vert\psi}$ and the noise should have zero mean when averaged over the map, we may subtract the pixel average, $\hat\kappa_{0} -\left<\hat \kappa_{0}\right>_{\rm pixel}$, to remove that zero-point function. Secondly, the estimator reconstructs the lensing contribution $\kappa_{0\vert\psi}$ with a calibration bias if non-Gaussian structures contribute to the lensing field \cite{amblard}. Although the CMB lensing efficiency peaks at redshift $z\sim 3,$ where most structure is linear, it has contributions from all redshifts so that inevitably it will be sourced by non-Gaussian structures.

The reconstructed maps depend on the CMB detector noise via the kernel in the estimator (see Ref.~\cite{carvalho}). The power spectrum of the CMB temperature detector noise 
depends on the characteristics of the simulated CMB experiment. For one frequency channel, the CMB noise power spectrum is given by \cite{knox95, tegmark97}
\ba
N(\ell)
=\left({\fwhm}{\sigpix}\right)^2
\exp[({\fwhm})^2\ell(\ell+1)/(8\ln2)].
\ea
Here $\fwhm$ is the beam full-width at half-maximum, which defines the pixel size in the simulated maps, and $\sigpix$ 
is the white noise amplitude per pixel in units $\mu$K, which depends on the detector sensitivity and observation time. Our maps simulate a fraction of the sky $\fsky$ smaller than the fraction of the sky covered by the target experiment $\fsky^{\rm fid}$. In order to have the same noise amplitude in the maps, we adopt the so-called active incomplete sky-coverage strategy  \cite{magueijo&hobson96}, which assumes the smaller coverage is compensated by longer observation times, reducing the noise per pixel as 
$\sigpix={\sigpix^{\rm fid}}\left(\fsky/\fsky^{\rm fid}\right)^{1/2}.$ Alternatively, this can be seen as scaling the pixel size such as to cover the fiducial fraction of the sky.

\subsection{The estimated power spectrum}

We turn now to the power spectrum of the convergence map. The noise in the map is assumed to be uncorrelated with the lensing signal, $\big<\kappa_{0\vert\psi}~\kappa^\ast_{0\vert\psi=0}\big>=0$, but it has a non-vanishing power spectrum $\big<\kappa_{0\vert\psi=0}~\kappa^\ast_{0\vert\psi=0}\big>$. Consequently, the convergence power spectrum computed from an unbiased convergence map is biased. This additive bias $N^{(0)}$ is evaluated in Ref.~\cite{kck} and can be subtracted off. Alternatively, we estimate convergence maps from various realizations of unlensed CMB maps for the same cosmology. The average of the power spectra over the different CMB realizations is an estimator of $\big<\kappa_{0\vert\psi=0}~\kappa^\ast_{0\vert\psi=0}\big>$. It also includes the additional bias in the case of a non-zero zero-point. We thus subtract it off of the estimated biased power spectrum to obtain a corrected estimator of the power spectrum
\ba
C_{\ell}^{\kappa_{0\vert \psi}\kappa_{0\vert \psi}}
=C_{\ell}^{\hat\kappa_{0}\hat\kappa_{0}}
-\left<C_{\ell}^{\hat\kappa_{0\vert\psi=0}\hat\kappa_{0\vert\psi=0}}\right>_{CMB}.
\label{eq:minusbias}
\ea

However, this estimator is still biased. Indeed, since $C_{\ell}^{\hat\kappa_{0}\hat\kappa_{0}}$ is a four-point correlator of the lensed CMB field, it also includes a connected part which introduces couplings between lensing modes, thus producing an additive bias $N^{(1)}$ which depends on the convergence power spectrum itself \cite{kck}. Further additive biases arise when expanding Eq.~(\ref{eq:cmblensing}) to higher orders in the lensing potential \cite{hanson10}. The magnitude of these higher-order biases was studied numerically in Ref.~\cite{amblard}.

\begin{figure}
\vskip-1.5cm
\centerline{
\includegraphics[width=12cm]
{
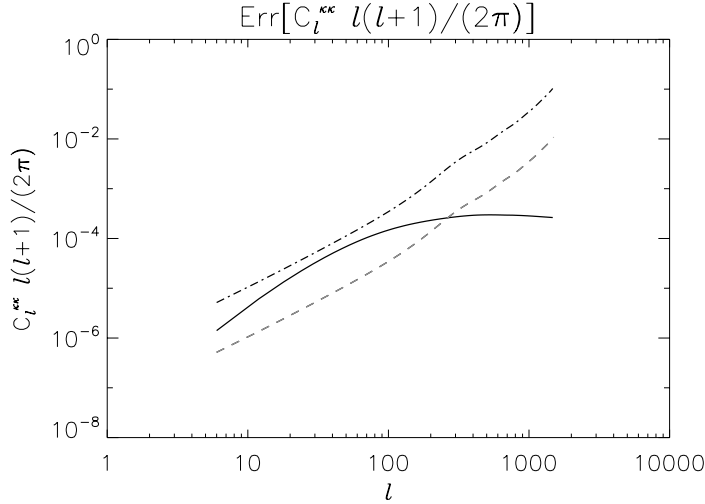}
}
\vskip-7.5cm
\caption{\baselineskip=0.5cm{ 
{\bf Error budget of the lensing power spectrum estimation for a \planck-like survey.} The three terms in the square-root of Eq.~(\ref{eq:varianceunbiased}) are shown in dashed-dotted (estimator noise), solid (cosmic variance), and dashed (bias uncertainty) lines.
}}
\label{fig:errbudget}
\end{figure}

Here we will use Eq.~(\ref{eq:minusbias}) to compute the convergence power spectrum, thus correcting for the lensing-independent noise-bias, and will also check for a multiplicative bias.  
We include three contributions to the variance of the corrected power spectrum, arising from: the variance of the convergence estimator (the noise), the cosmic variance and the uncertainty on the removed bias. 
The theoretical variance of the estimator per binned mode ${\rm Var}[\hat \kappa_{0}]$ is derived in Ref.~\cite{carvalho}, Appendix A. Adding the cosmic variance contribution yields the variance of the biased estimator of the power spectrum. The uncertainty on the additive bias is the variance of the mean 
$\big<C_{\ell}^{\hat\kappa_{0\vert\psi=0}\hat\kappa_{0\vert\psi=0}}\big>_{CMB},$ which was estimated by averaging the convergence power spectra over $N$ realizations of unlensed maps. It is given by
\ba
{\rm Var}[C_{\ell}^{\kappa_{0\vert\psi=0}\kappa_{0\vert\psi=0}}]=
\left(\sum_{i=1}^N\,\frac{1}{{\rm Var}[C_{\ell}^{\kappa_{0\vert\psi=0}\kappa_{0\vert\psi=0}}]_i}
\right)^{-1},
\ea
where, for each map, the variance ${\rm Var}[C_{\ell}^{\kappa_{0\vert\psi=0}\kappa_{0\vert\psi=0}}]_i$ is computed as in the first term of Eq.~(\ref{eq:varianceunbiased}).
The variance of the corrected power spectrum is obtained by adding the contributions in quadrature, written in terms of variance per band as follows
\ba
{\rm Var}[C_{\ell}^{\kappa_{0\vert \psi}\kappa_{0\vert \psi}}]
\approx{1\over {\ell~\fsky}}
\left( {\rm Var}[\hat \kappa_{0}]
+C_{\ell}^{\kappa_{0\vert\psi}\kappa_{0\vert\psi}}\right)^2+
{\rm Var}[C_{\ell}^{\kappa_{0\vert\psi=0}\kappa_{0\vert\psi=0}}].
\label{eq:varianceunbiased}
\ea
The denominator accounts for the number of modes probing the multipole band, which depends on the area of the simulated maps. To forecast the error budget of a given experiment, its fiducial fraction of the sky must be used instead. The variance-covariance matrix is essentially diagonal \cite{kck}, i.e. the multipoles may be considered independent.

The three contributions are shown in Fig.~\ref{fig:errbudget} for a \planck-like survey with the noise specifications of Tab.~\ref{table:ena} and $\fsky=1$. The noise term dominates on all scales. The bias is removed using 100 realizations, which is a large-enough sample to produce a sub-dominant bias uncertainty.

\section{Application to masked lensed maps}
\label{sec:application}

\subsection{The point source mask}

We generate the map
\ba
\tilde M=[MB+N][SB],
\label{eq:mask}
\ea
where M, B, N and S refer respectively to lensed CMB, beam, noise and point source maps, as described below.

We start by generating a lensed CMB map using the formalism described in Ref.~\cite{carvalho}, Appendix B. In summary, we first compute the TT and the $\psi\psi$ power spectra generated by CAMB \cite{camb}, on scales $\ell\le 4000$, for a fiducial flat $\Lambda$CDM cosmology consisting of $h=0.7,\, \Omega_{\rm cdm}h^2=0.112,\, \Omega_{\rm b}h^2=0.0226,\, n_{\rm s}=0.96,\, \sigma_8=0.85,\, \tau_{\rm rei}=0.09,\, T_{0}=2.725\,\rm{K}$.
With these power spectra, we build a CMB temperature map and a deflection map on a square patch. We then apply the deflection map to the CMB map to produce a lensed CMB map. This operation consists in shifting the CMB map by the lensing potential assuming the Born approximation.
We then introduce the detector effects, first by convolving the lensed map with the beam profile and then by adding a map constructed from the detector noise power spectrum.
The size of a map is given by the pixel size times the number of pixels, 
with the pixel size being determined by the beam size.
We use two cases based on the experimental characteristics of \planck, denoted by $Designer$ and $Planck$, the first containing no detector noise, the second containing that of the $\nu=143~{\rm GHz}$ \planck\, channel of the HFI instrument \cite{planck_hfi}.
Finally, we mask the map by placing disks of finite diameter at the point source locations, where the amplitude is set to zero.

\begin{table}[t]
\begin{tabular}{cccccccc}
\hline
Experiment ~&~$\nu$
~&~$\fwhm$ ~&~$\sigpix^{\rm fid}$ & $\fsky^{\rm fid}$ & ${\rm FOV_{map}}$ 
~&~$\ns$~&~$\theta_{\rm s}$~\\ 
&(GHz)&$(\rm arcmin)$ &$(\mu K)$&
&$({\rm deg^2})$~&$({\rm deg^{-2}})$&$(\rm arcmin)$\\
\hline
$Designer$~~
&143& 7.2 &0 & 1& 61.3$\times$61.3 &0.03&36\\ 
$Planck$~~
&143& 7.2 &7.5 & 1& 61.3$\times$61.3 &0.03&36\\ \hline
\end{tabular}
\caption{\label{table:ena} \baselineskip=0.5cm{
{\bf The specifications of the two experiments used in this study.} The experiment denoted by $Planck$ is inspired by the $\nu=143~{\rm GHz}$ channel of the {\planck} HFI instrument \cite{planck_hfi}. The experiment denoted by $Designer$ is the same as $Planck$ for no detector noise.
}}
\end{table}
 
The mask is generated by placing a number density $\ns$ of excisions
on locations of the map randomly generated with a uniform distribution.
The value of $\ns$ used for the cases of $Planck$ and $Designer$ is obtained from the total number of point sources detected on the full sky at the 143 GHz channel, as reported in the {\planck} early results \cite{planck_ps}. 
The mask should be convolved with the beam, which is equivalent to assigning to each point source a disk of finite diameter $\theta_{\rm s}$. The size of the masking disk $\theta_{\rm s},$ assumed uniform, must be proportional to the size of the beam $\fwhm$. Indeed, a higher-resolution experiment, i.e. one with a smaller beam, would smear the flux of each source over a smaller area and in addition it would detect more point sources. 
Hence, if the sources are uniformly distributed, the product $\ns\theta_{\rm s}$ is constant across experiments. 
For example, the analysis of the Atacama Cosmology Project (ACT) 2008 Survey used masks with  $\theta_{\rm s}=5\,\fwhm$ \cite{marriage}, for the $\ns=0.35 \deg^{-2}$ point sources detected with a beam $\fwhm \approx 2 \,{\rm arcmin}$ \cite{act_sources}. This implies masking disks of the same relative size,  $\theta_{\rm s}=5\,\fwhm$, for \planck\,.
Thus, in our masks we will use for the masking disk diameter the value $\theta_{\rm s}=36\,{\rm arcmin}.$ We note that pixels where masks overlap count as one point source only. 

\subsection{The convergence of a masked lensed map}

We apply the real-space estimator to a lensed CMB map with and without the point source masks, and obtain estimated maps of the convergence which we denote respectively by $\tilde\kappa_{0}$ and $\kappa_{0}$ (whenever there is no case for confusion, to simplify the notation we drop the hat that identifies an estimated quantity). For comparison, we also implement the harmonic-space estimator on the same maps (with the implementation also described in \cite{carvalho}). 
In addition, we run several realizations of unlensed CMB maps. Each realization consists of a different map built from the same temperature power spectrum. For each unlensed CMB map, we apply various realizations of the mask with the same $\ns$ and $\theta_{\rm s}$ but with the excisions located at different positions.  

\begin{figure}[t]
\setlength{\unitlength}{1cm}
\centerline{
\hskip2.3cm
\includegraphics[width=9.5cm]
{
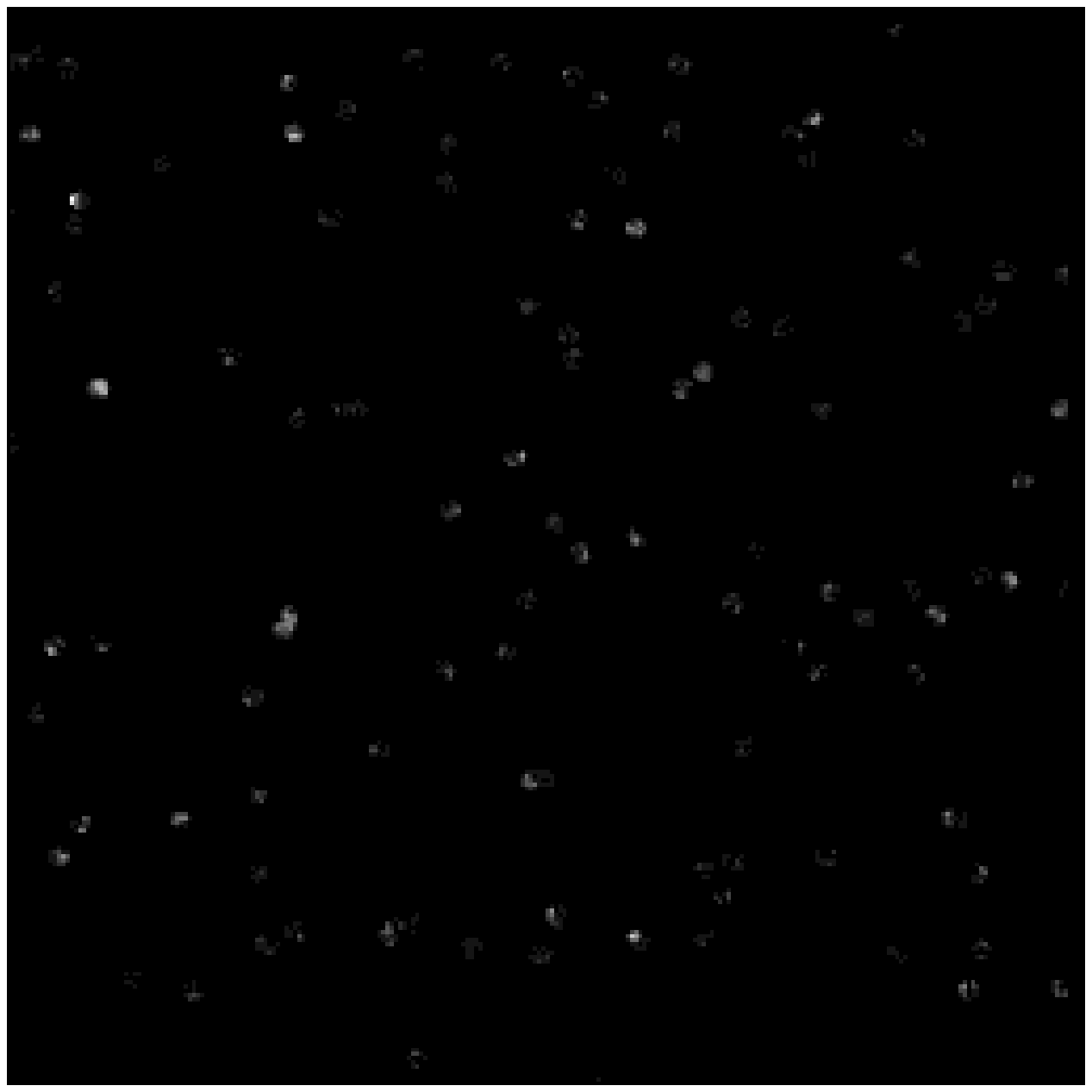}
\hskip-3.9cm
\includegraphics[width=9.5cm]
{
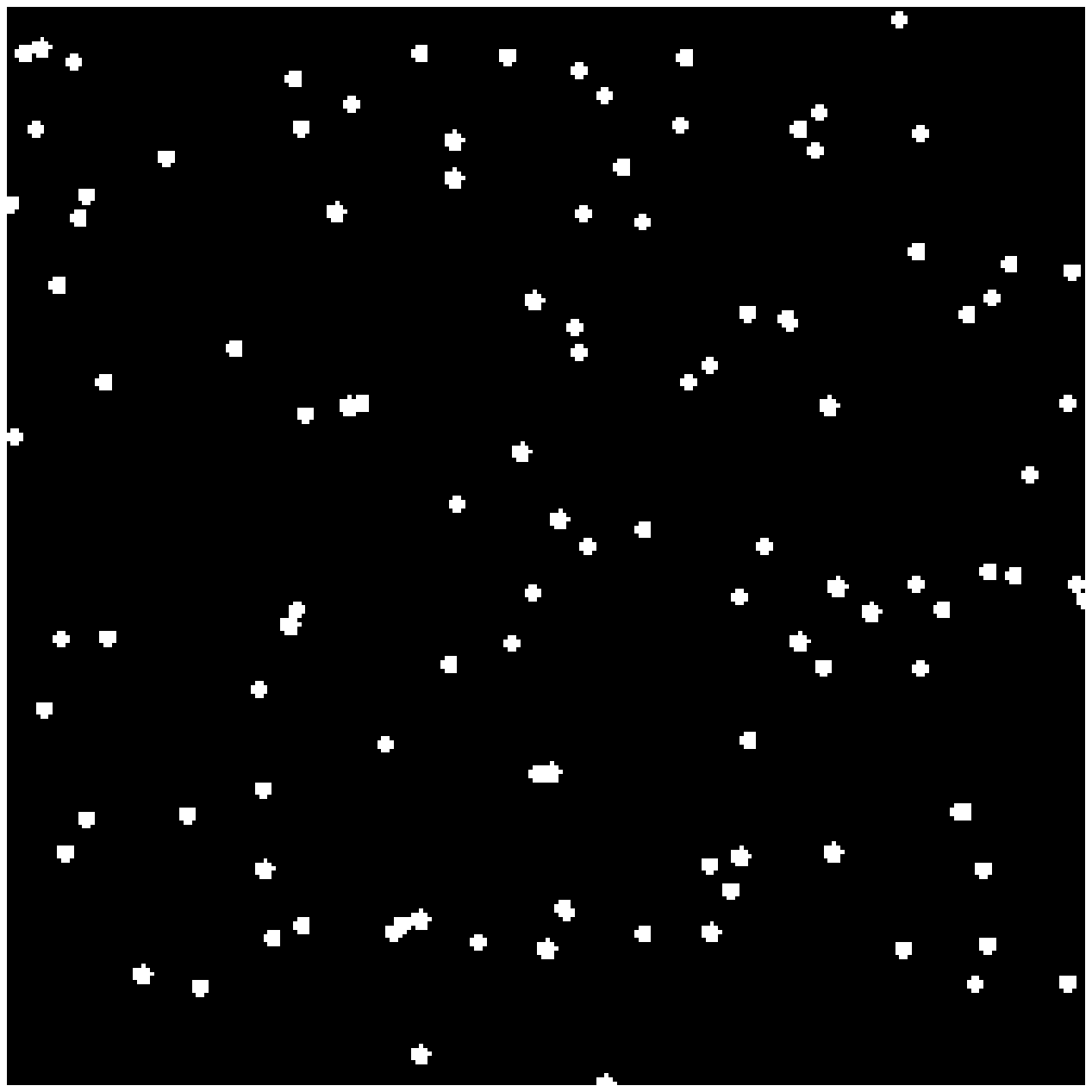}
\hskip-3.9cm
\includegraphics[width=9.5cm]
{
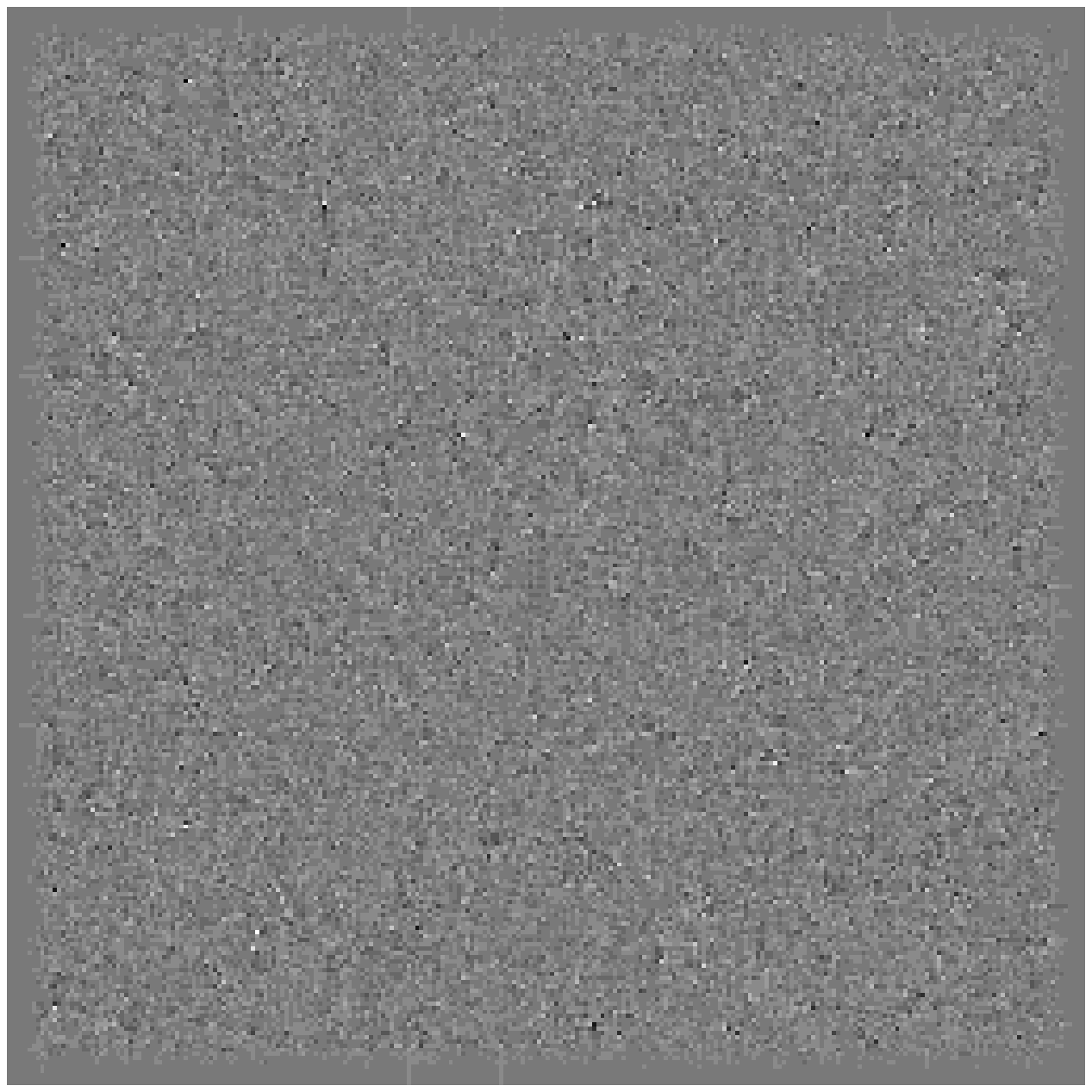}
}
\vskip-5cm
\caption{\baselineskip=0.5cm{ 
{\bf Maps for the $Planck$ experiment.} 
Left panel: the difference $(\kappa_{0}-\tilde\kappa_{0})$ between the two maps reconstructed with the real space estimator is shown in a greyscale ranging from black (no difference) to white (largest difference).
Central panel: the mask map, with amplitude 0 on the white dots representing the effective size of the point sources, and amplitude 1 elsewhere. 
Right panel: the $(\kappa_{0}-\tilde\kappa_{0})$ map obtained from the harmonic space estimator, same scale as in the left panel.}}
\label{fig:mask_map}
\end{figure}

\begin{figure}[t]
\setlength{\unitlength}{1cm}
\vskip-1.5cm
\centerline{
\hskip-0.5cm
\includegraphics[width=10cm]
{
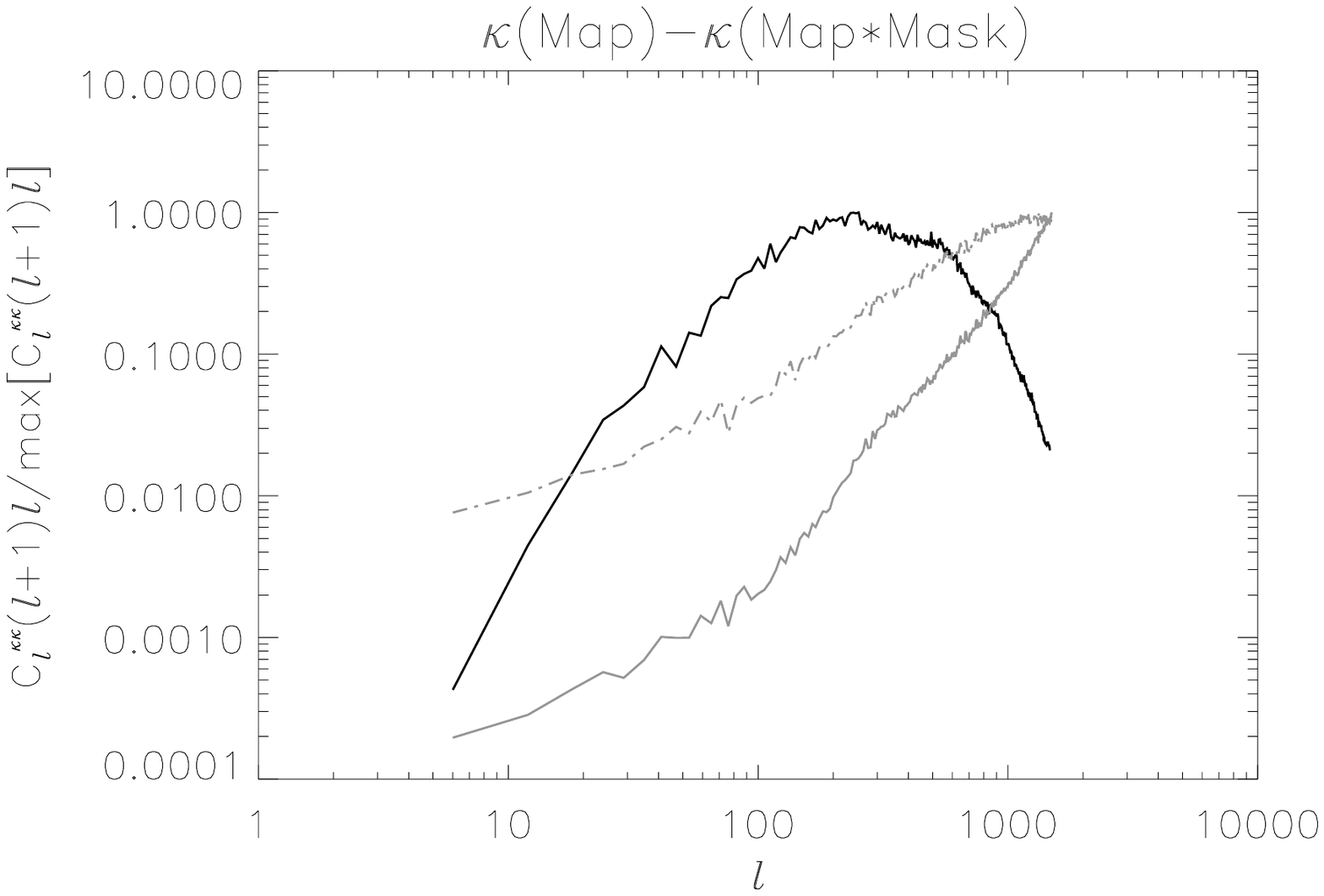}
\hskip-1.5cm
\includegraphics[width=10cm]
{
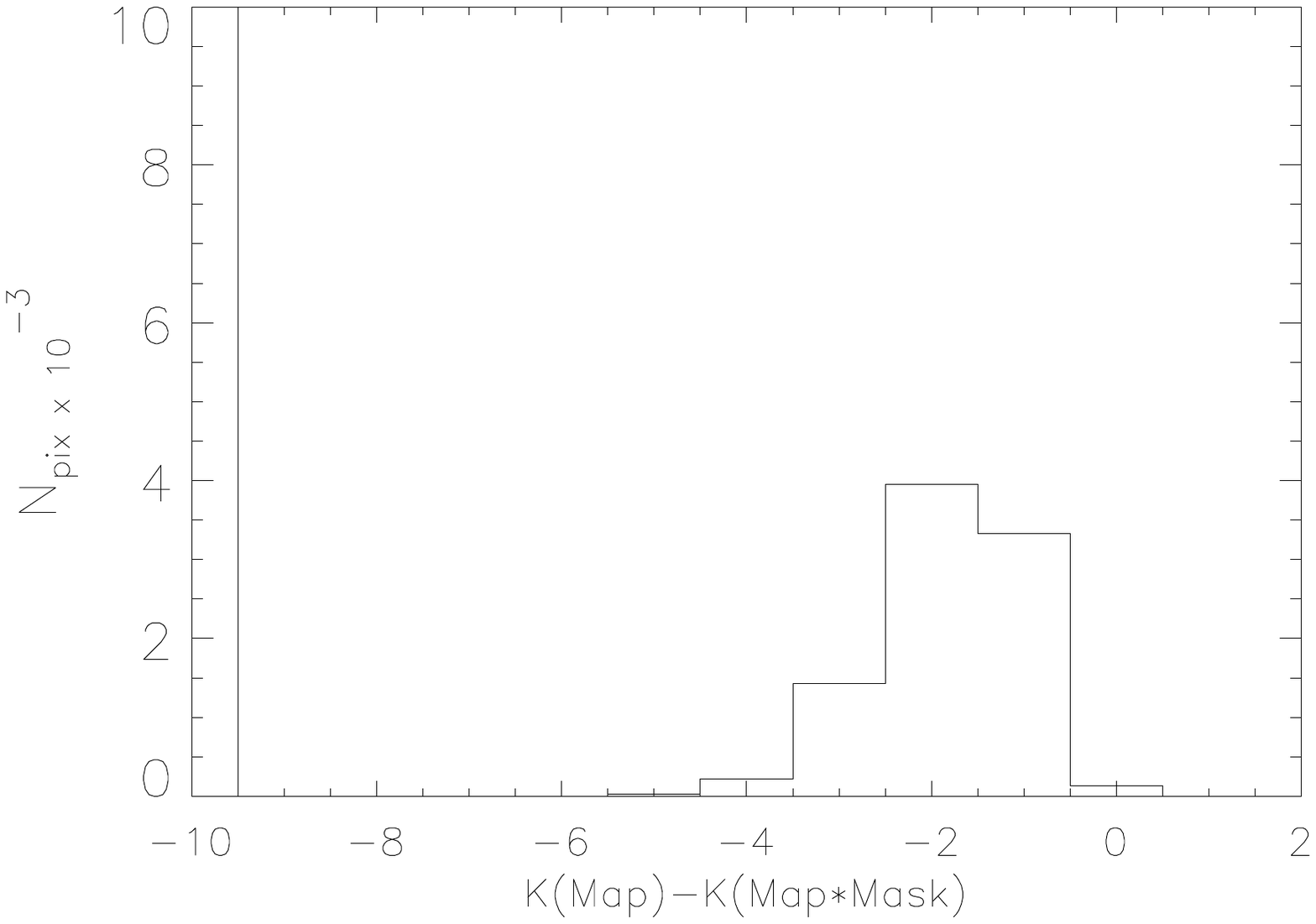}
}
\vskip-6 cm
\caption{\baselineskip=0.5cm{ 
{\bf 
The $(\kappa_{0}-\tilde \kappa_{0})$ map.} 
Left panel~: The power spectrum of the difference map for the real space (black) and harmonic space (gray: dashed-dotted for $Designer,$ full for $Planck$) estimators, with arbitrary amplitude. Right panel~: The histogram of the difference map obtained with the real space estimator.
}}
\label{fig:diff}
\end{figure}

The reconstruction made with the real-space estimator for a masked map, $\tilde \kappa_{0}$, is very similar to the reconstruction $\kappa_{0}$, only missing a small amount of the structure obtained in $\kappa_{0}$. The difference between the two maps $\kappa_{0}-\tilde \kappa_{0}$ (shown in Fig.~\ref{fig:mask_map}, left panel) is confined to localized medium-scale structure similar in scale and location to that of the mask used in this realization (shown for comparison in Fig.~\ref{fig:mask_map}, central panel). 
For the case of the harmonic-space estimator the result is very different, as shown in  Fig.~\ref{fig:mask_map} (right panel). Indeed, due to the highly nonlocal character of the harmonic filter, there is a spread of artifacts from the disks over the whole map, producing a large amount of new structure which overwhelms the lensing structure.
This is alternatively seen in the power spectrum of the difference maps, shown in Fig.~\ref{fig:diff} (left panel). The power spectrum of the difference map produced with the harmonic-space estimator without detector noise increases at all scales. Adding detector noise, this increase of power is predominant at small scales. In contrast, the power spectrum of the difference map in the real-space case is peaked, showing an increased difference on scales  $\ell \sim 200,$ which corresponds to roughly three times the size of the masking disks, and implying a localized small spread of the artifacts. The histogram in Fig.~\ref{fig:diff} (right panel) shows that $97\%$ of the pixels have an identical value in $\tilde \kappa_{0}$ and $\kappa_{0}$. The remaining $3\%$, which corresponds to three times the area masked according to Tab.~\ref{table:ena}, are distinctly grouped at values lower than the typical amplitude of a noisy pixel $\sigma_{\kappa_0}=0.1$. 

We note that the comparison of the power spectra shapes in Fig.~\ref{fig:diff} (left panel) is made with arbitrary amplitudes. This is because the normalization in the estimators fails in the presence of noise and masking, and the maps become affected by different calibrations. The multiplicative bias of a map can be evaluated by taking the ratio of the cross power spectrum $C_{\ell}^{\hat \kappa \kappa_{in}}$ by the auto power spectrum of the input convergence map $C_{\ell}^{\kappa_{in}\kappa_{in}}.$ Assuming that the noise in the reconstructed power spectrum is uncorrelated with the lensing information, this ratio is a direct measure of the calibration bias. 

\subsection{The convergence power spectrum of a masked lensed map}

\begin{figure}[t]
\setlength{\unitlength}{1cm}
\vskip-1.5cm
\centerline{
\hskip-0.5cm
\includegraphics[width=12cm]
{
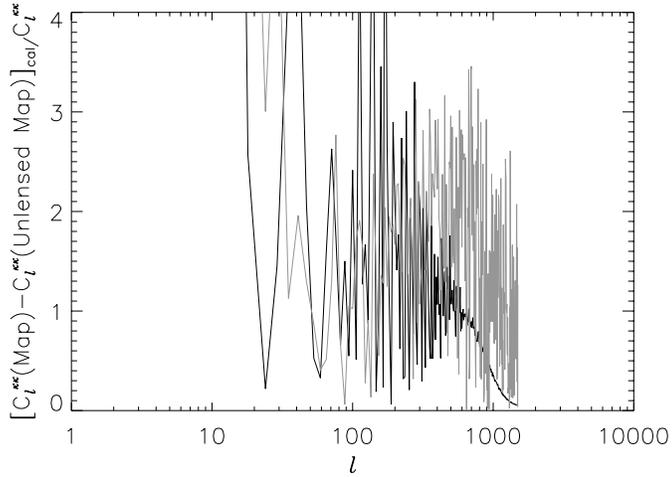}
}
\vskip-7.5cm
\caption{\baselineskip=0.5cm{ 
{\bf The multiplicative bias.} The residual multiplicative bias in the power spectra for the real-space (black line) and harmonic-space (gray lines) estimators. 
}}
\label{fig:designer_bias}
\end{figure}

We compute now the convergence power spectrum and its variance from the convergence maps estimated from lensed CMB maps, both in the case of masked and unmasked maps and for the two experiments of Tab.~\ref{table:ena}. The power spectra are computed for all $\ell$-modes in the range $6 <\ell<1500$. The small-scale limit corresponds to twice the pixel size, according to the Nyquist sampling theorem, and the large scales are limited by the size of the maps. We follow the procedure described in Sec.~\ref{sec:estimator}, subtracting the mean convergence power spectrum from several realizations of the unlensed sky, as described by Eq.~(\ref{eq:minusbias}), which succeeds in removing most of the additive bias. We then apply a calibration derived from the ratio of the estimated power spectrum (after subtracting the additive bias) 
to the input power spectrum. In particular, we use the case $Designer$ as the control experiment for calibration, and average this ratio over the region where the two power spectra are approximately parallel i.e. $70<\ell<400.$ After this correction, we remain with the residual multiplicative bias shown in Fig.~\ref{fig:designer_bias}. The residual, shown with no binning, fluctuates around 1 in the range $70<\ell<400,$ with the fluctuations being larger in the case of the harmonic-space estimator.  

\begin{figure}[t]
\setlength{\unitlength}{1cm}
\vskip-1.5cm
\centerline{
\hskip-0.5cm
\includegraphics[width=12cm]
{
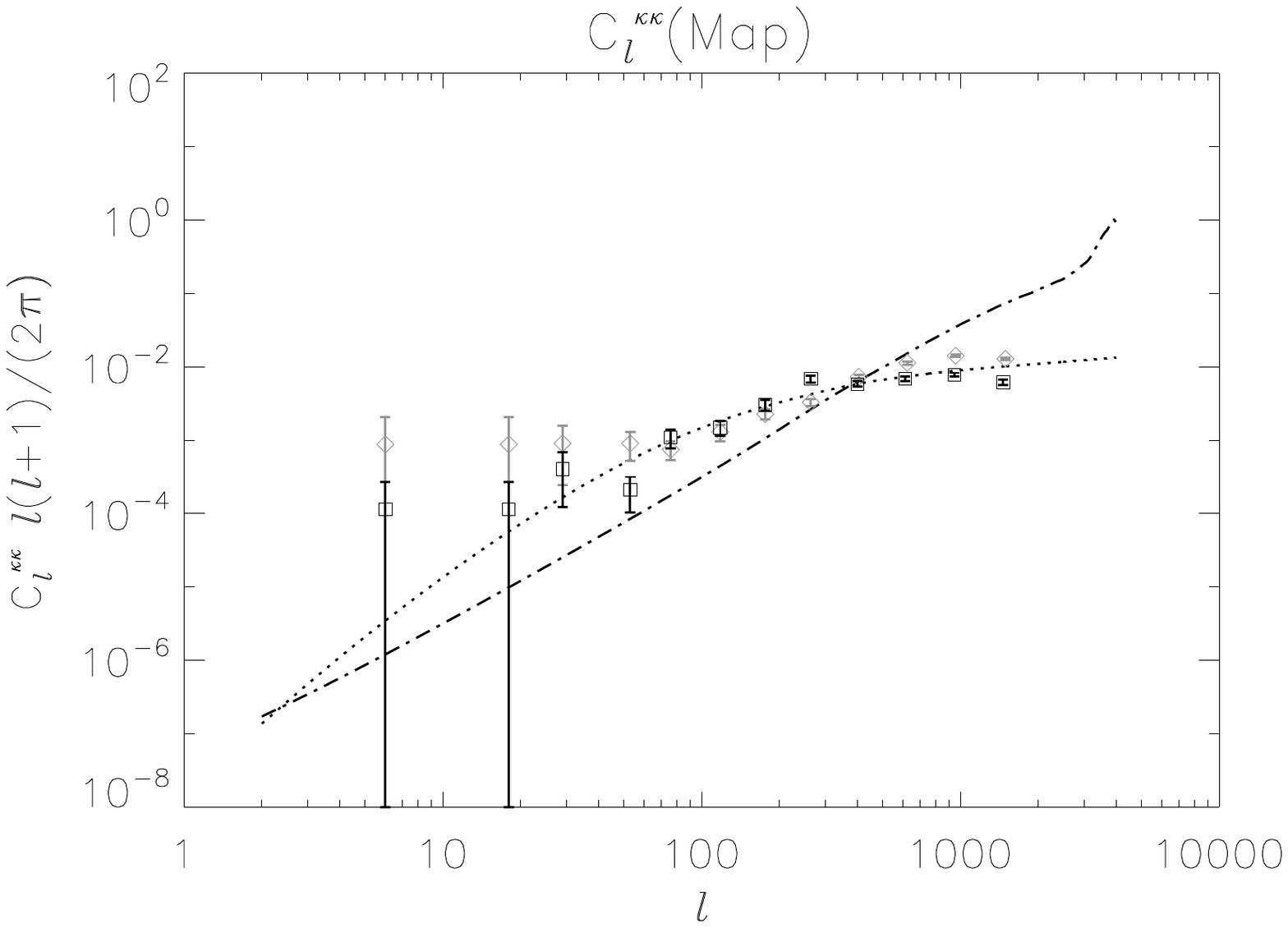}
\hskip-3cm
\includegraphics[width=12cm]
{
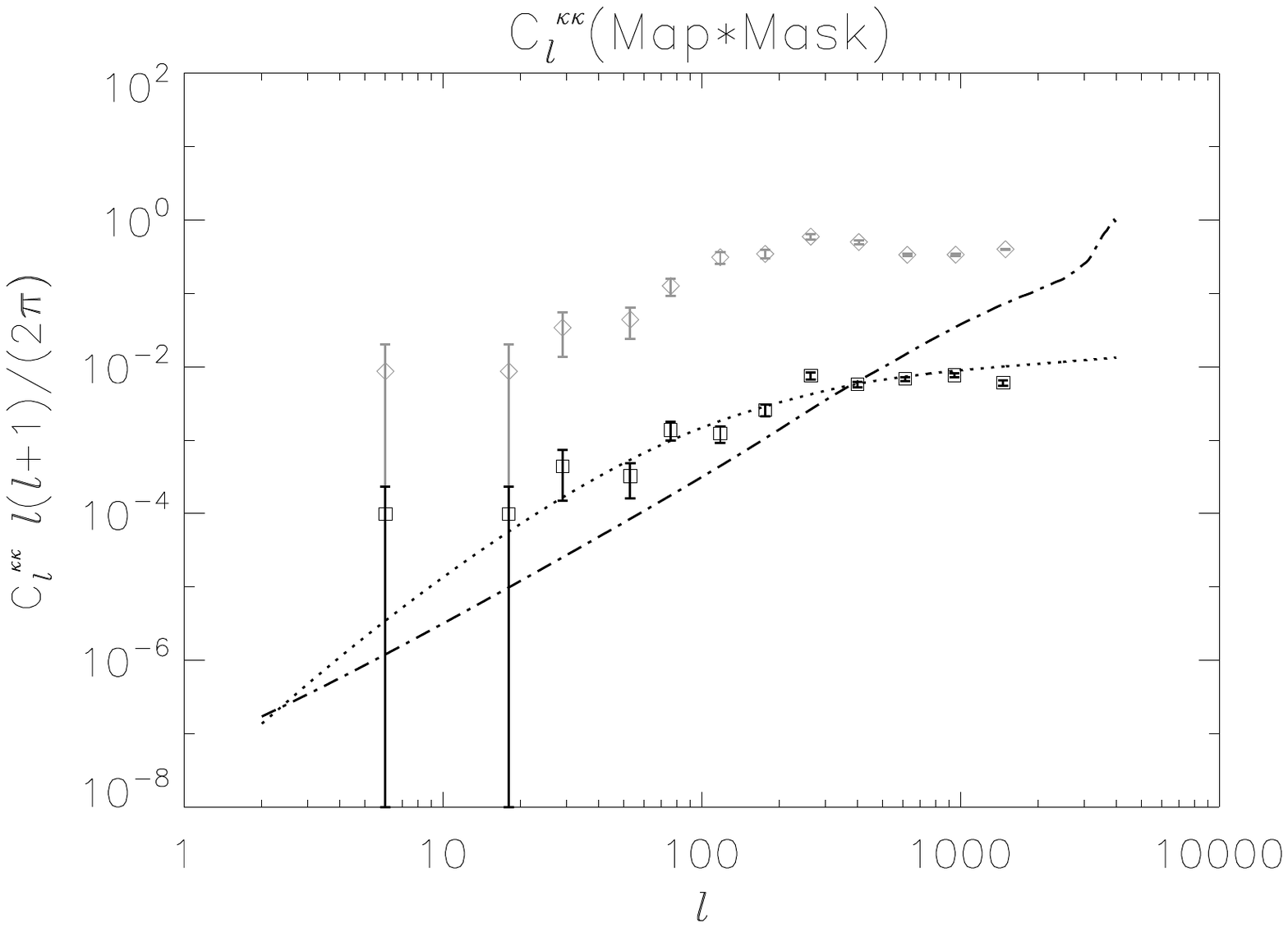}
}
\vskip-7.5cm
\caption{\baselineskip=0.5cm{ 
{\bf 
Estimated convergence power spectrum for $\designer$.} Left panel~: Pristine lensed map. Right panel~: Masked lensed map.
In both cases are shown the reconstruction with the harmonic-space estimator (gray diamonds), with the real-space estimator (black squares), the input power spectrum $C^{\kappa_{in}\kappa_{in}}$ (dotted), and the estimator noise for $\fsky^{\rm fid}.$ 
}}
\label{fig:cl_kk2_designer_planck}
\end{figure}

\begin{figure}[t]
\setlength{\unitlength}{1cm}
\vskip-1.5cm
\centerline{
\hskip-0.5cm
\includegraphics[width=12cm]
{
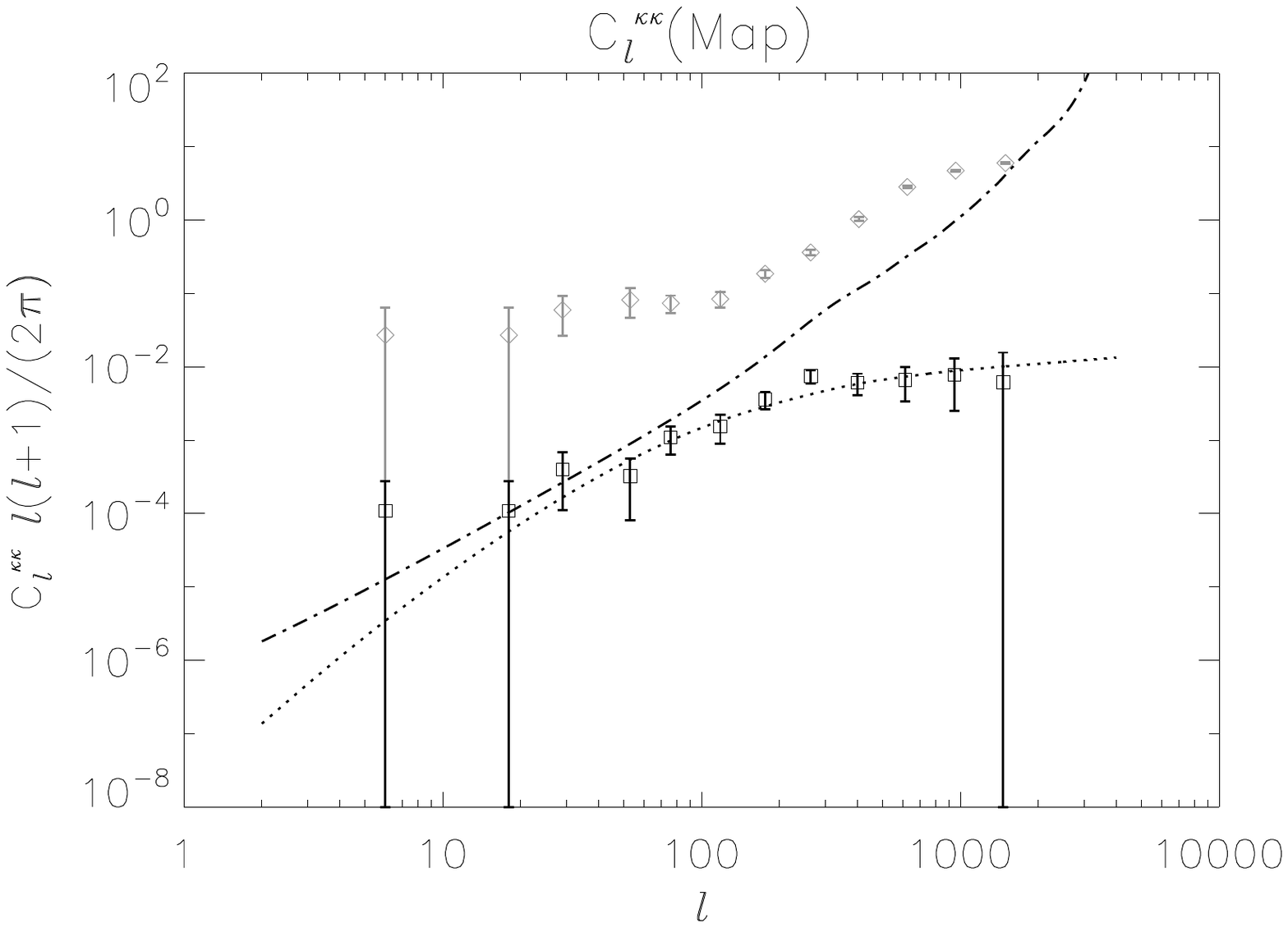}
\hskip-3cm
\includegraphics[width=12cm]
{
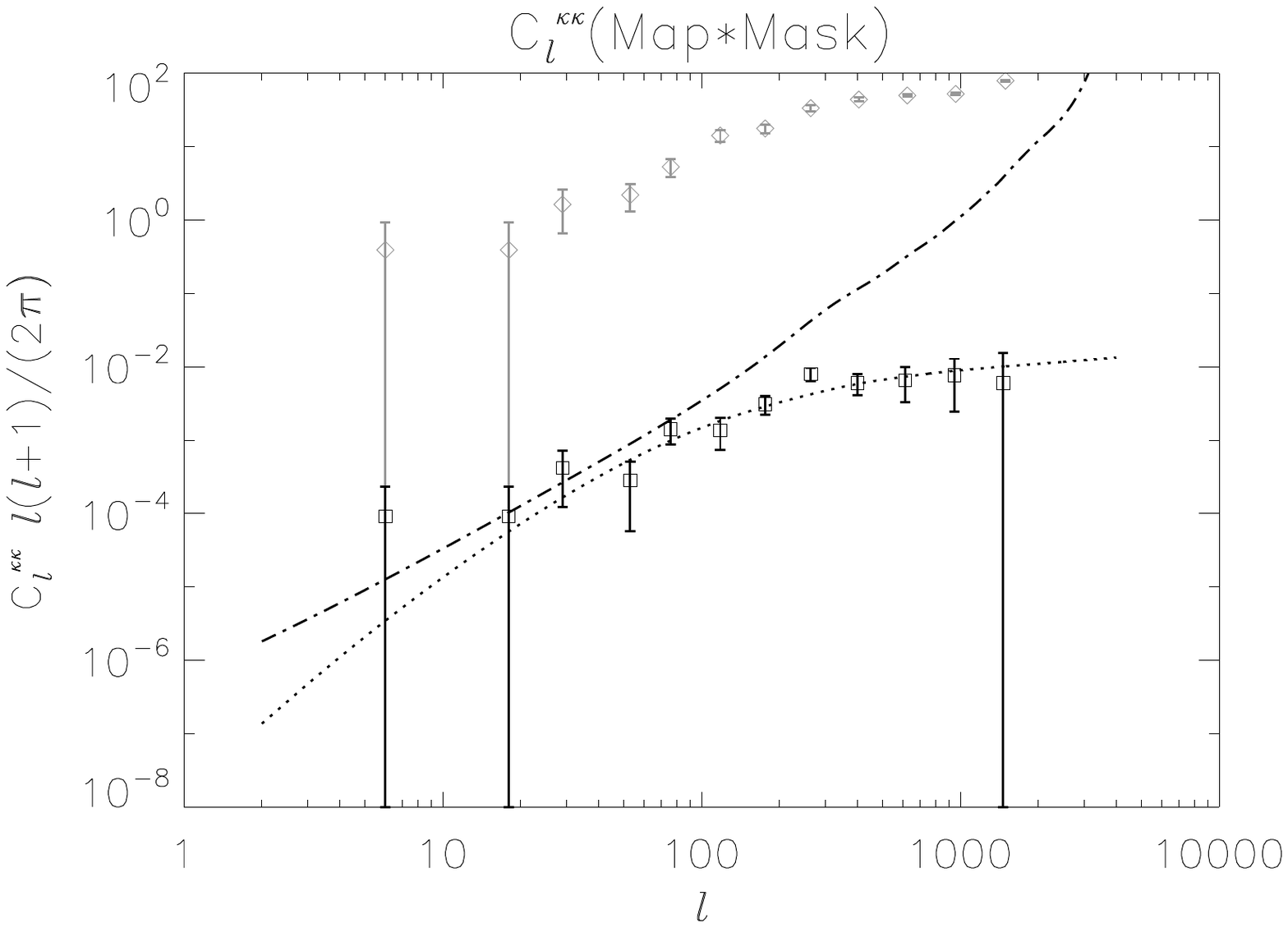}
}
\vskip-7.5cm
\caption{\baselineskip=0.5cm{ 
{\bf 
Estimated convergence power spectrum for $\planck$.} Left panel~: Pristine lensed map. Right panel~: Masked lensed map.
In both cases are shown the reconstruction with the harmonic-space estimator (gray diamonds), with the real-space estimator (black squares), the input power spectrum $C^{\kappa_{in}\kappa_{in}}$ (dotted), and the estimator noise for $\fsky^{\rm fid}.$ 
}}
\label{fig:cl_kk2_planck}
\end{figure}

The results for the debiased and calibrated convergence power spectrum are shown in Fig.~\ref{fig:cl_kk2_designer_planck} for $Designer$ and in Fig.~\ref{fig:cl_kk2_planck} for $Planck.$ 
The power spectrum is binned in logarithmically spaced intervals and the error bars are correspondingly reduced according to the number of $\ell$-modes per bin. 
The error bars include the effective detector noise per pixel, as discussed earlier on, and are produced for the effective size of the maps. Forecasted error bars for the \planck\, survey are smaller by a factor of 3 and the power spectrum of the estimator noise for that case (assuming $\fsky^{\rm fid}=1$) is also shown.

The real-space estimator produces similar results for all the four cases (whether masked or unmasked, with or without detector noise). 
The four power spectra obtained show similar shapes and amplitudes and the same calibration is applied to all.
In particular, the reconstruction on the masked maps is remakably similar to the reconstrucion on the unmasked maps, just slightly 
lower around $\ell \sim 200$ due to loss of power in the masked regions seen in Fig.~\ref{fig:diff} (left panel). 
Concerning the reconstruction on maps in the presence of detector noise, the estimator is insensitive to uncorrelated noise, since the sum of the product of pairs of neighbouring pixels, weighted by the kernel, averages out uncorrelated pixel noise. 
Also notice that for all cases the recovered power on the smallest scales is limited by the size of the kernel $(\ell = 1000)$, visible as a decrease in the last two bins. Indeed, as discussed in Ref.~\cite{carvalho}, there is an averaging of the modes smaller than the extent of the kernel.

The behaviour of the harmonic-space estimator is very different. It performs well only in the case of no detector noise and no masking. In the presence of detector noise, it picks up the white noise, as it was shown in Ref.~\cite{carvalho}, increasing the power on the noise-dominated small scales. 
In the presence of masking, the spurious power changes both the shape and amplitude of the estimated power spectrum, and the calibration found with the pristine map no longer applies.

\section{Varying the mask parameters}
\label{sec:degradation}

We have two attributes to characterize the point source mask: the number density of sources and the masking area per source.
To understand how these two parameters affect the lensing reconstruction, 
we produced maps where we either a) varied the number density of excisions while keeping the diameter of the masking disks constant at $\theta_{\rm s}=5\,\fwhm,$ or b) varied the diameter while keeping the number density constant at $n_{\rm s}$. 

\begin{figure}[t]
\setlength{\unitlength}{1cm}
\vskip-1.5cm
\centerline{
\hskip-0.5cm
\includegraphics[width=12cm]
{
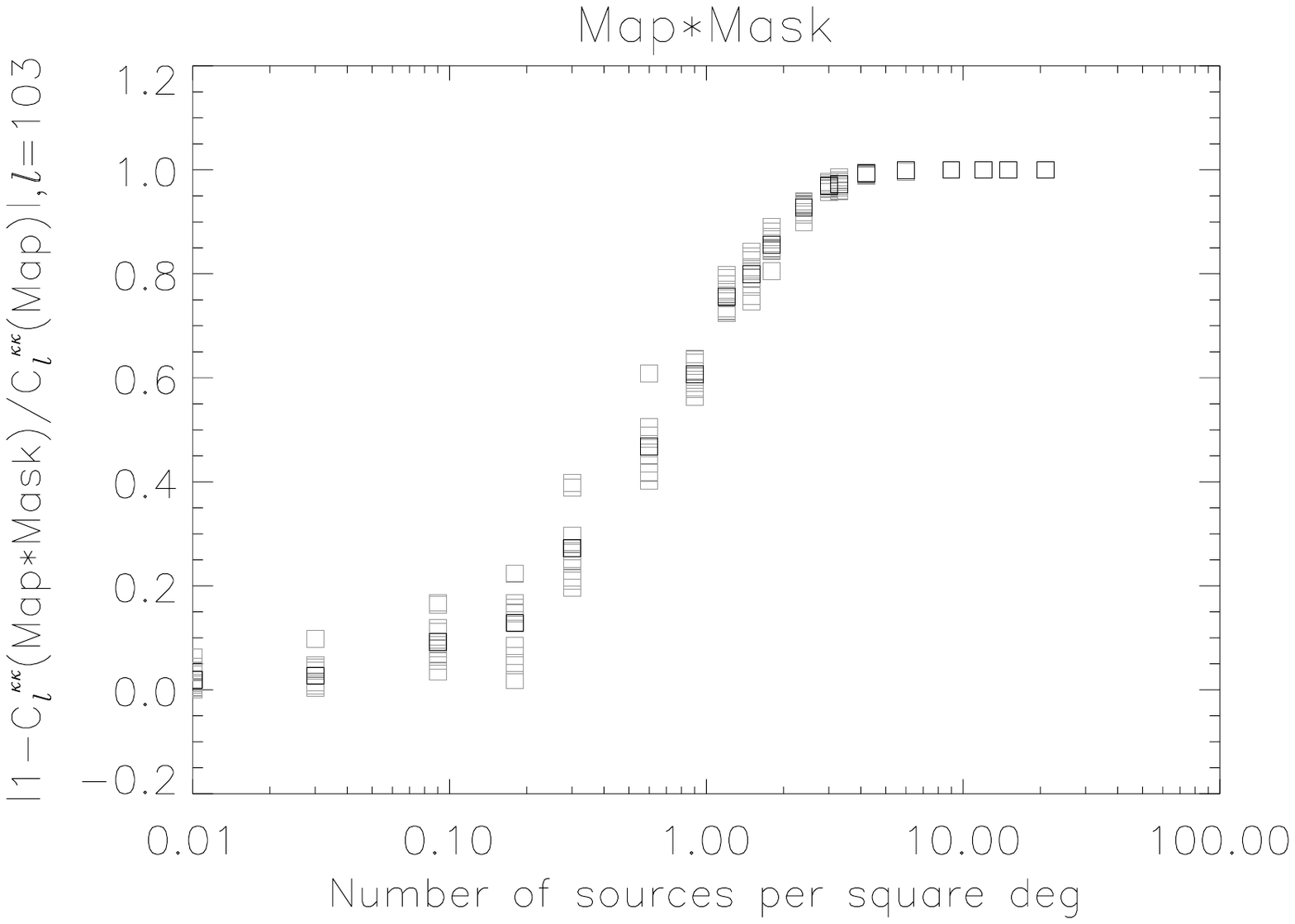}
\hskip-3cm
\includegraphics[width=12cm]
{
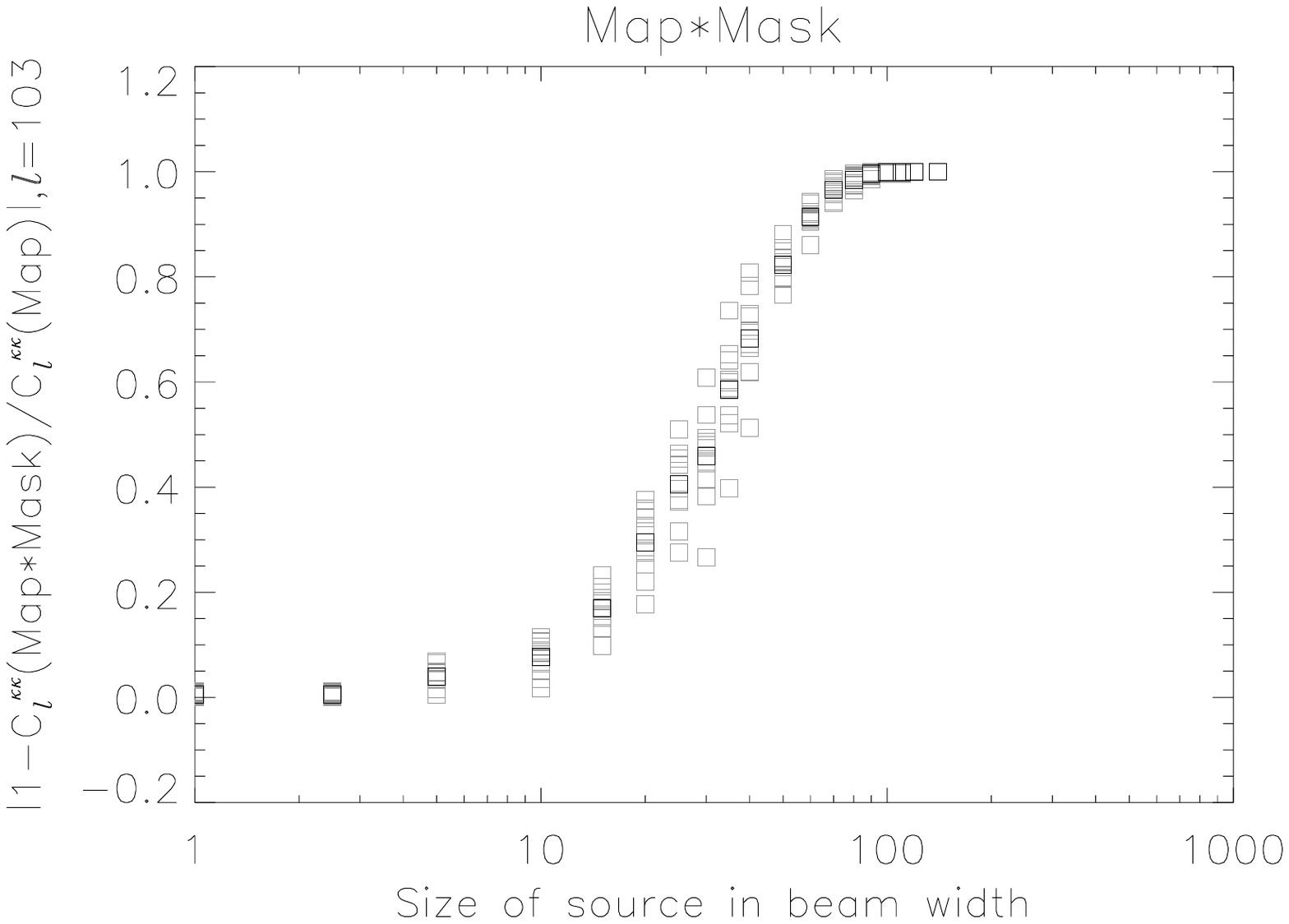}
}
\vskip-7.5cm
\caption{\baselineskip=0.5cm{ 
{\bf Degradation of the reconstructed convergence power spectrum with the mask parameters.}
Left panel~: The normalized difference between the power spectrum modes $\ell=103$ reconstructed with and without masking with the real-space estimator for the case of $\planck$, are shown as a function of $n_{\rm s}$ for each of the 10 realizations (gray squares) and averaged over the 10 realizations (black squares). Right panel~: As the left panel, but as a function of $\theta_{\rm s}$. 
}}
\label{fig:cl_kk2_mapXmask_dens}
\end{figure}


We start by assessing the effect of the number density of masked sources in the estimation of the convergence.
We produce maps with $\theta_{s}=5\,\fwhm$ for
$n_{\rm s}\in\{ 
0,0.01,0.03,0.09,0.18,0.3,0.6,0.9,1.2,1.5,1.8,2.4,3.0,3.3,4.2,6.0,9.0,12,15,21 
\}\,{\rm deg}^{-2},$
which corresponds to 
$N_{\rm s}\in\{ 
0,38,113,338,676,1126,2252,3379,4505,5631,6757,9009, 11262,\\
12388,15767,22524,33785,45047,56309,78833
\}$ sources. 
For each value of the number density, we generate ten realizations of the mask which we then apply to one realization of the lensed CMB map. We implement the real-space estimator to the resulting masked maps, and present the results in Fig.~\ref{fig:cl_kk2_mapXmask_dens} (left panel) showing the fractional difference between the power spectrum estimated with each of the masks and the unmasked case. The difference is normalised with the power spectrum estimated for the unmasked case, and is shown for the mode $\ell=103$. The fiducial case (for \planck) is the point at $n_{\rm s}=0.03~{\rm deg}^{-2}.$

We observe that the estimation of the convergence power spectrum from a map which has been masked degrades monotonically with the increase of the density of excisions in the mask as compared to the estimation from the complete map. Both the $\planck$ and $\designer$ (not shown) cases show similar behaviours, with the reconstructions in the presence of noise detector showing a larger dispersion.
The amplitude of the power spectrum in the presence of masking decreases with the masking density due to the loss of power discussed in Sec.~\ref{sec:application}.
In particular, in a $(x,\ln[y])$ plot, $\ln[C_{\ell=103}^{\hat \kappa_{0}}]$ is a negative linear function of the excision number density $n_{s},$ which implies that the degradation may be written in the form $C_{\ell=103}^{\hat \kappa_{0}}(n_{\rm s})=C_{\ell=103}^{\hat \kappa_{0}}(0)\exp[-\alpha\,n_{\rm s}].$ 
The monotonically decrease of $\ln[C_{\ell=103}^{\hat \kappa_{0}}]$ implies that the degradation tends to $100\%,$ which is reached when the mask covers the whole map. 


We proceed to assess the effect of the size of the masking disks in the estimation of the convergence.
We generate point source masks with different masking disk diameters $\theta_{\rm s}$ 
while keeping the number density constant. In particular, we use
$n_{\rm s}=0.03~{\rm deg}^{-2},$ which corresponds to 113 sources, for
$\theta_{\rm s}\in\{
0,1, 2.5, 5, 10, 15, 20, 25, 30, 35, 40, 50, 60, 70, 80, 90, 100, 110, 120, 140
\}\,\fwhm.$
For each value of the diameter, we generate ten realizations of the point source mask which we then apply to the same realization of the lensed CMB map. 
We implement the real-space estimator to the resulting masked and unmasked maps, and show the resulting fractional deviations in Fig.~\ref{fig:cl_kk2_mapXmask_dens} (right panel), where the fiducial case (for \planck) is the point at $\theta_{\rm s}=5.$

As before, we observe that the estimation of the convergence from a map which has been masked degrades with the size of the masking disks as compared to the estimation from the complete map.
In a  $(x,\ln[y])$ plot, $\ln[C_{\ell=103}^{\hat\kappa_{0}}]$ is now a negative quadratic function of the disk diameter $\theta_{\rm s},$ which implies that it can be written as $C_{\ell=103}^{\hat \kappa_{0}}(\theta_{\rm s})= C_{\ell=103}^{\hat\kappa_{0}}(0)\exp[-\theta_{\rm s}^2/(2\sigma^2)].$  

\begin{figure}[t]
\setlength{\unitlength}{1cm}
\vskip-1.5cm
\centerline{
\hskip-0.5cm
\includegraphics[width=12cm]
{
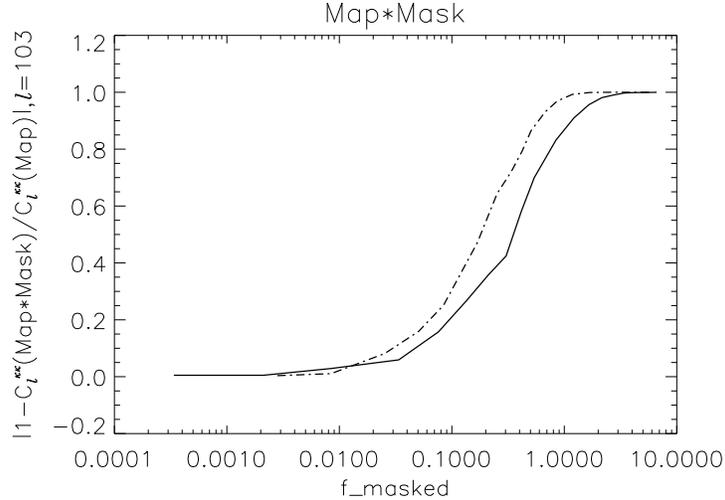}
}
\vskip-7.5cm
\caption{\baselineskip=0.5cm{ 
{\bf Amplitude degradation.}
Normalized difference between the power spectrum modes $\ell=103$ reconstructed with and without masking with the real-space estimator as a function of the masked fraction, for varying $n_{\rm s}$ (solid) or varying $\theta_{\rm s}$ (dashed-dotted).
}} 
\label{fig:fmask}
\end{figure}

\begin{figure}[t]
\setlength{\unitlength}{1cm}
\vskip-1.5cm
\centerline{
\hskip-0.5cm
\includegraphics[width=10cm]
{
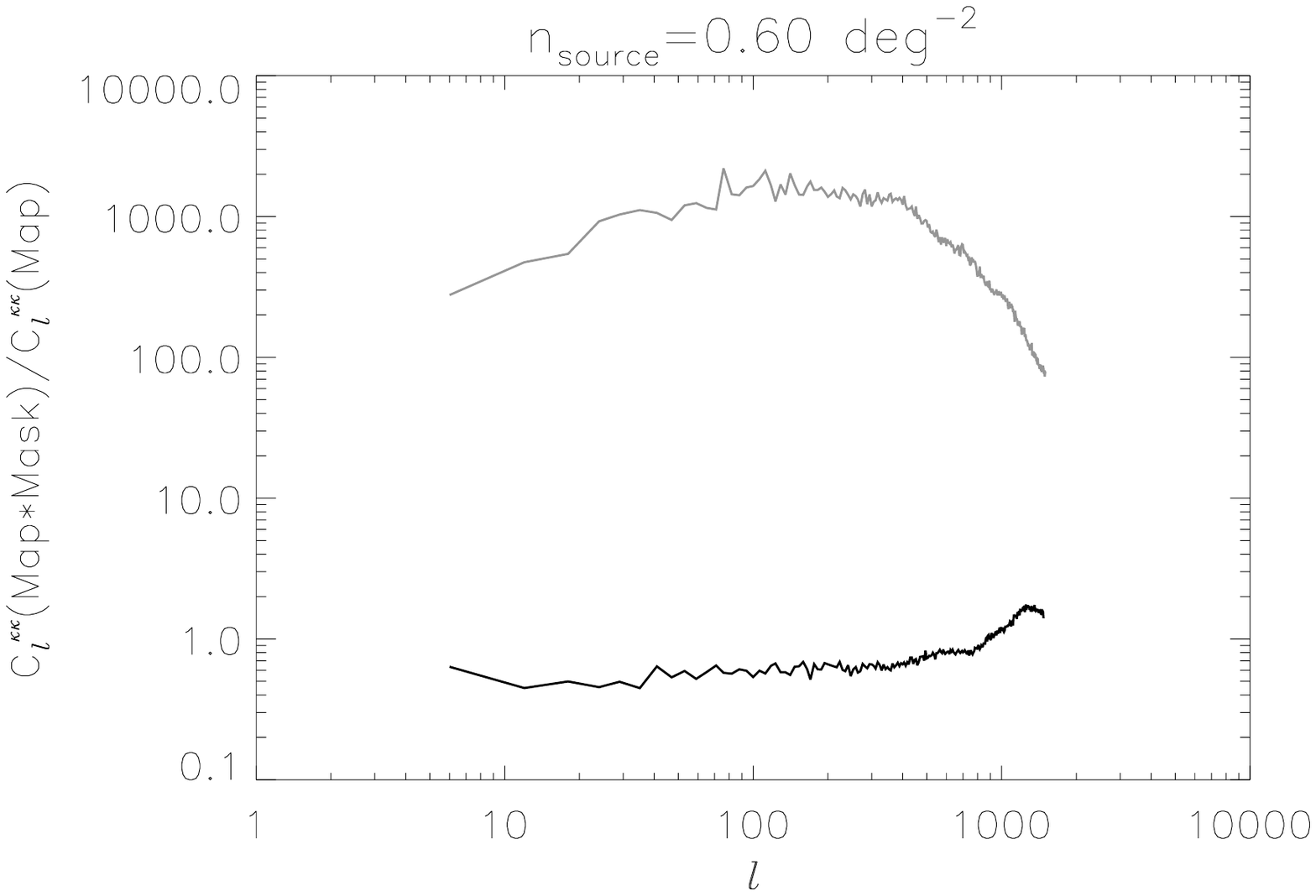}
\hskip-2cm
\includegraphics[width=10cm]
{
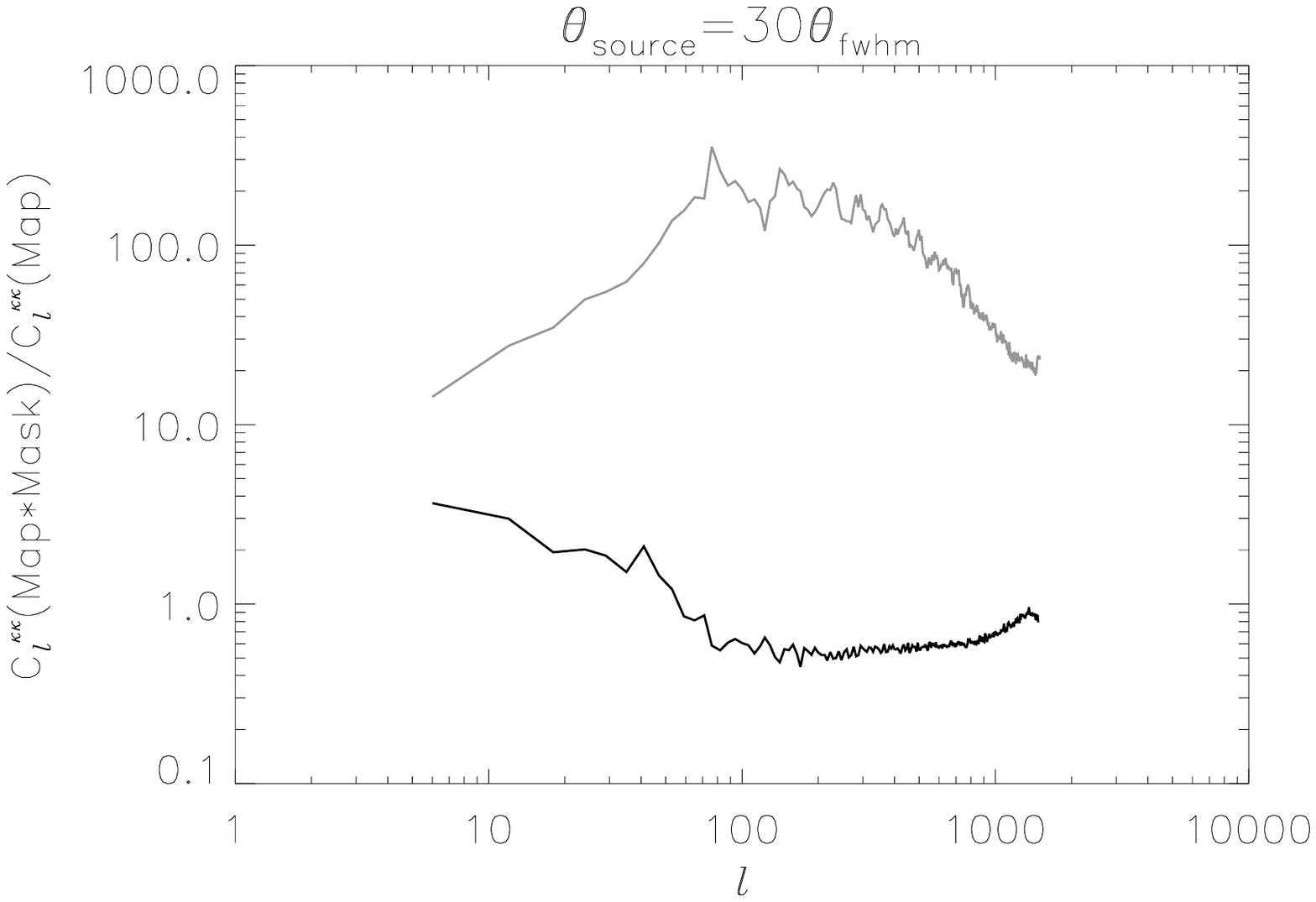}
}
\vskip-6.5cm
\caption{\baselineskip=0.5cm{ 
{\bf Shape degradation.} Left panel: Ratio between the power spectra with and without masking reconstructed with the real-space (black line) and harmonic-space (gray line) estimators, for a mask of fiducial size and $n_{\rm s}=0.6\,{\rm deg}^{-2}$ density. Right panel: The same for a mask of fiducial density and size $\theta{\rm s}=30\,\fwhm.$
}}
\label{fig:densdiam}
\end{figure}

We can directly compare the two analyses if we translate $n_{\rm s}$ and $\theta_{\rm s}$ into masked fractions.
Fig.~\ref{fig:densdiam} shows this comparison in terms of the power losses on the scale $\ell=103$.
We observe that the amplitude degrades faster with the density, i.e. for the same masked area, the loss of power is greater when more disks are placed than when the area of the disks is increased. The two cases have different geometries, in the first case the disks cover the map more uniformly, while in the second case larger contiguous unmasked areas are possible.

Besides the change in amplitude in an individual mode, we also compare the two analyses in terms of their impact on the shape of the power spectrum. For this we consider two cases where the loss of amplitude in $C_{\ell=103}^{\kappa}$ amounts to about $50\%$. These are: a) $n_{\rm s}=0.60\,{\rm deg}^{-2}$ for the fiducial $\theta_{\rm s},$ corresponding to a masked area  $f_{\rm masked}\approx 0.2$ of the map, and b) $\theta_{s}=30\,\fwhm$ for the fiducial $n_{\rm s},$ corresponding to $f_{\rm masked}\approx 0.3.$ The plot in Fig.~\ref{fig:densdiam} (left panel) shows that the loss of power is roughly scale-independent. In contrast, the variation of the size of the disks tilts the power spectrum shifting power to large scales (Fig.~\ref{fig:densdiam}, right panel). The scales where the power increases are larger than the size of the enlarged reconstructed disk discussed in Sec.~\ref{sec:application}, which for this example corresponds to $\ell \sim 40$.  The shift of power is observed until $\theta_{\rm s}\sim 40\,\fwhm,$ beyond which the overall decrease in power overwhelms this tilt.

We also applied the harmonic-space estimator to these tests, even though it is outside the goal of our analysis to proper calibrate this estimator in the presence of masks. We found an increase of power of one to three orders of magnitude throughout all angular scales and spurious oscillations (Fig.~\ref{fig:densdiam}).
When the masked area reaches $7\%$ ($26\%$) for the case of varying $n_{\rm s}$ ($\theta_{\rm s}$), the power spectrum flattens and its amplitude starts to decrease with the masked area, throughout all scales. 

\section{Summary}
\label{sec:summary}

In this manuscript we study the effect of point source masks in the reconstruction of the weak lensing convergence from CMB maps. We used a variant of the minimum variance estimator with truncated support in real space. For comparison, we also used the original minimum variance estimator with full support in harmonic space.

First we compared the reconstructions of convergence maps and power spectra made in complete versus masked temperature maps.
The number of excisions in the masked map was determined by the number density of point sources reported in the literature, while the diameter of the masking disks was suggested by the size adopted in the ACT survey for modelling point sources.

When using the real-space estimator, we found no difference in the quality of the estimation in the presence of masking.
The input amplitude and shape were recovered after subtracting off the additive bias and applying a calibration vector derived from the reconstruction made in the absence of noise and masking.

Conversely, when using the harmonic-space estimator, we found a decrease in the quality of the estimation of the convergence of around $40\%$ at the map reconstruction level. We found that the amplitude and shape of the input power spectrum could not be recovered in the presence of noise or masking without further treatment of the calibration, since the simple method of applying a calibration vector derived from the reconstruction in the absence of noise and masking cannot work in the presence of mode-coupling. There are several methods in the literature to deal with the problem of missing data. For example in Ref.~\cite{hivon} a method was developed to remove masking effects using a mode-coupling matrix. The ``full" power spectrum is then calculated from the reconstructed ``pseudo" power spectrum by means of a deconvolution involving matrix inversion, or with Bayesian methods. This was used in the recent first direct detection of a lensing potential in CMB data \cite{smidt10}. Another method consists in pre-whitening the map by applying an inverse covariance matrix weighting, in order to have a map that is less affected by mode-coupling. Such method was used in the first indirect detection of gravitational lensing in the CMB \cite{ksmith07}.

We then proceeded to analyse the degradation of the reconstruction of the convergence power spectrum as a function of the parameters of the mask, varying the masked area between 0 and $100\%$.

We observe that the real space estimator yields less power as more map area is masked, with the power spectrum being typically recovered at $80\%$ when the masked area reaches $10\%$ of the map. As a subdominant effect, we observe a shift of power from small to large scales, which is a consequence of the smearing of the masked disks observed in the histogram of the difference map.

These observations strongly favour the use of the real-space estimator for the estimation of the weak lensing convergence in the case of maps with excision of points. The robustness of the real-space estimator, whether we vary the number density or the size of the masking disks, suggests that it can be used reliably on maps where extended areas have been masked. 

The computational cost of the real-space implementation scales with the number of pixels $N^2$ on the CMB map and inversely with the pixel size $\theta_{\rm fwhm},$ whereas that of the harmonic-space implementation scales with $N\log[N].$ In particular, in the fiducial experiment used here, where $N=512$ pixels and $\theta_{\rm fwhm}=7.2\,{\rm arcmin},$ the real-space estimator takes twice as long as our implementation of the harmonic-space estimator. Doubling the resolution, the real-space estimator takes four times longer.

The problem of missing data is particularly relevant for the extraction of the polarization signal from the CMB, where it prevents a perfect E and B component separation \cite{bunn02}.
Further studies can be envisaged to test the potentialities of this approach for lensing with CMB polarization.

\vspace{0.6cm}
\acknowledgments
CSC is supported by Funda\c{c}\~ao para a Ci\^encia e a Tecnologia (FCT), Ref.~SFRH/ BPD/ 65993/ 2009.
IT is funded by FCT (SFRH/BPD/65122/2009) and acknowledges support from the European Programme FP7-PEOPLE-2010-RG-268312.
The authors thank M~Hilton, T~Marriage and K~Moodley for useful discussions. 
\hfill

\end{document}